# On Theoretical Stability Proof and Stability Margin Analysis of Enhanced Droop-Free Control Schemes for Islanded Microgrids

Weipeng Liu, *Member, IEEE,* Upendra Prasad, Yutian Liu, *Senior Member, IEEE,* Yong Dong, Haoran Zhao, *Senior Member, IEEE,*
Lei Wu, *Fellow, IEEE*

*Abstract*—This paper studies enhanced droop-free control strategies with sparse neighboring communication for achieving effective active power sharing of distributed energy resources (DERs) while maintaining the frequency stability of islanded microgrids. The normalized active power consensus (NAPC) based droop-free control can share the load among controllable DERs in proportion to their available capacities. However, existing literature exclusively takes the asymptotic stability of the NAPC-based droop-free control for granted, lacking a comprehensive theoretical proof that is critical for ensuring its effective design and practical implementation. This paper, for the first time, provides a thorough theoretical proof of the asymptotic stability of two NAPC-based droop-free control schemes: ordinary NAPC (O-NAPC) and amplifier-equipped NAPC (A-NAPC), by testifying that all effective eigenvalues have negative real parts. The effect of various system settings on the stability margins is further analyzed with respect to the average admittance of the electrical network, the sparseness of the communication network, and the average available capacity of controllable DERs. Based on the sensitivity of eigenvalues with respect to perturbations, a vulnerability analysis is conducted to identify the weaknesses in the microgrids. Case studies demonstrate that the available capacity of controllable DERs has the most decisive influence on the stability margin of NAPC-based droop-free control, while O-NAPC/A-NAPC control scheme is more suitable for microgrids with DERs of larger/ smaller available capacities.

## I. INTRODUCTION

The increasing penetration of distributed energy resources (DERs) such as solar power imposes huge challenges to the supply-demand balance of power systems [1]. Recent frequent natural disasters and extreme weather events have exposed the inherent vulnerability of power systems, leading to high risks of widespread outages and severe economic losses. Benefiting from a large fleet of energy storage resources (ESRs), the energy efficiency and resiliency of microgrids could be greatly augmented [2, 3]. With this, microgrids have attracted significant interest because of their potential capabilities in enhancing power balance and energy resilience for end users [4]. Specifically, islanded microgrids, equipped with comprehensive control schemes to maintain the active and reactive power balance and provide frequency and voltage support against disturbances, can operate in a self-sustained manner without relying on the bulk power grid.

The hierarchical control structure has been widely applied to islanded microgrids, by activating primary, secondary, and tertiary controls at different time scales to imitate the frequency and voltage regulation processes of bulk power systems [5, 6]. Specifically, the primary control responds promptly to disturbances [7], the secondary control recovers frequency and voltage back to the nominal values in approximately a few seconds [8], and the tertiary control is triggered every few minutes to update the active and reactive power set points for achieving optimal operations [9]. This paper focuses on the frequency and active power control to regulate nodal frequencies at the nominal value against disturbances, realizing the combined functionalities of primary and secondary control in the hierarchical control structure.

Indeed, within the hierarchical control structure, primary control is usually realized via a decentralized droop method on local voltage source converters of DERs to imitate the dynamics of synchronous generators, and secondary control is deployed on either a microgrid central controller (MGCC) or distributed local controllers to coordinate DERs [10, 11]. However, this hierarchical control structure suffers from inherent issues: (i) Droop control is sensitive to measurement errors and thus requires sensors of high quality [12]; (ii) MGCC induces a high risk of single-point failure that will collapse the microgrid system [13], while maintaining a reliable MGCC and the associated communication network for islanded microgrids is expensive. Besides, the computational burden of MGCC could also hinder the scalability of such microgrid systems; and (iii) For the distributed secondary control, the control system is complicated and the optimal parameter setting becomes challenging [14].

Different from the droop-based hierarchical control, distributed droop-free control was studied in [12, 13, 15-20] to achieve zero frequency deviation at the steady state without the need for the MGCC. With the droop-free control scheme, local disturbances are first balanced by local droop-free controlled DERs; Then, multiple droop-free controllers are coordinated to achieve proper active power sharing and zero frequency deviation through a sparse neighboring communication network [14]. Major benefits of droop-free control include: (i) the averaged nodal frequency remains at the nominal value throughout the entire dynamic process, and (ii) it realizes the functionalities of both primary and secondary controls in the droop-based hierarchical control.

Based on the information exchanged among DERs, droop-free control can reach different active power-sharing consensus. Specifically, if DERs exchange their active power outputs with their neighbors, the steady state expresses that all controllable DERs are at the same active power level, i.e., active power consensus (APC) [13, 15-17]; if DERs exchange normalized active power information with their neighbors, the steady state expresses that all controllable DERs are at the same normalized active power level, i.e., normalized active power consensus (NAPC) [18-20].

The APC-based droop-free control for microgrids has been extensively studied, and its asymptotic stability property has been well understood. Specifically, reference [13] proved that all eigenvalues of the microgrid with APC are non-positive, and the system becomes unstable if the electrical network or communication network is separated. Reference [15] further proved that all the system's effective eigenvalues are negative if the power and communication networks are symmetric and connected graphs. Lyapunov functions were also used in [17] to prove the stability of APC-based droop-free control.

As it is more proper to adjust the active power outputs of DERs in proportion to their available capacities, NAPC-based droop-free control was further studied. However, existing studies [18-20] exclusively take the asymptotic stability of the NAPC-based droop-free control for granted, lacking a comprehensive theoretical understanding. The steady-state performance of microgrids with NAPC was analyzed in [18, 19], showing that the steady-state is a stationary point of the system. The small-signal stability of microgrids with NAPC was investigated [19, 20] by applying the Prony analysis to facilitate control parameter settings and avoid undesired oscillation. However, in these literatures, the system stability is verified only under a few specific application scenarios or control structures. The theoretical understanding of the stability of the NAPC control scheme in general settings remains undiscovered.

The above literature review clearly shows that the asymptotic stability of NAPC-based droop-free control for microgrids has not undergone comprehensive theoretical verification yet, which is critical for ensuring its proper design and practical implementation. To this end, for the first time, this paper provides a thorough theoretical proof of the asymptotic stability of two NAPC-based droop-free control schemes: ordinary NAPC (O-NAPC) and amplifier-equipped NAPC (A-NAPC), for microgrids with symmetric neighboring communication networks. Stability margins of the two control schemes with respect to various settings on the electrical and communication networks are further analyzed, which can help augment stability margins by selecting proper control schemes based on microgrid specifications.

The main contributions are summarized as follows:

(i) Along with the O-NAPC based droop-free control, the A-NAPC control scheme is further studied, especially for enhancing the stability of droop-free controlled microgrids with controllable DERs of smaller capacities.

(ii) It is the first time, to the best of our knowledge, that a thorough theoretical proof of the asymptotic stability of O-NAPC and A-NAPC designs is provided, by testifying that all effective eigenvalues have negative real parts.

(iii) The stability margins of the two droop-free control schemes are analyzed comparatively by the sensitivity analysis on various system parameter settings and the vulnerability analysis on perturbations of the system matrix.

The remainder of the paper is organized as follows. Section II presents the O-NAPC and A-NAPC designs for droop-free controlled microgrids. The theoretical proofs of the asymptotic stability of O-NAPC and A-NAPC designs are discussed in Section III, followed by the quantification of their stability margins. Case studies are conducted in Section IV to validate the two droop-free control schemes through hardware-in-the-loop (HIL) tests and demonstrate the efficacy of the vulnerability analysis. The paper is concluded in Section V.

## II. FORMULATION OF DROOP-FREE CONTROLLED MICROGRIDS

This section describes the state-space models of droop-free controlled microgrids with O-NAPC and A-NAPC designs. The following two assumptions [15] are adopted, enabling us to concentrate on frequency and active power control. Nevertheless, based on our previous study [15], the proposed O-NAPC and A-NAPC droop-free control designs can be extended to encompass broader scenarios (i.e., with non-zero $R/X$ ratios and filter dynamics) without compromising the stability performance.

*Assumption 1*: By leveraging the linearized power flow model and frame transformation [21], the active and reactive power flows can be decoupled, enabling the independent designs of frequency and voltage controls. With this, we focus on the frequency and active power control by using the susceptance matrix to model the electrical network.

*Assumption 2*: Filters, as a key component to provide clean signals, are assumed well-designed and will not compromise the controllers' dynamic performance. The generalized model including the formulation of the filter system was discussed in our previous work [15], demonstrating that the filter dynamics will not destabilize the system. With this, filters are not included in the microgrid models for the sake of discussion.

*Assumption 3*: For the sake of discussion, the electrical network formulation in this paper is presented in a way that every node is associated with a controllable DER. The generalized electrical network with certain nodes only associated with uncontrollable DERs and/or loads can be equivalently transformed to an electrical network that has a controllable DER connected to each node. The equivalent network transformation is shown in the Appendix.

Based on *Assumption 1*, active power flows can be modeled in (1) [13]. $\boldsymbol{p}_{net}(t)$ is the vector of nodal net active power injections, consisting of active power of controllable DERs (e.g., ESRs) and uncontrollable resources (e.g., renewable resources and loads); $\boldsymbol{B}$ is the $N$-dimensional susceptance matrix of the electrical network; $\boldsymbol{\theta}(t)$ is the vector of nodal phase angles. The dynamics of nodal phase angles can be presented by nodal frequencies $\boldsymbol{\omega}(t)$ as (2), where $\dot{\boldsymbol{\theta}}(t)$ is the derivative of $\boldsymbol{\theta}(t)$, $\omega_0$ is the nominal angular frequency, and $\mathbf{1}_N$ is a 1-vector of size $N$.

$$\boldsymbol{p}_{net}(t) = \boldsymbol{B} \cdot \boldsymbol{\theta}(t) \tag{1}$$

$$\dot{\boldsymbol{\theta}}(t) = \boldsymbol{\omega}(t) - \omega_0 \cdot \mathbf{1}_N \tag{2}$$

### A. The O-NAPC Based Droop-Free Control

The O-NAPC based droop-free controller is presented in (3) -(5) to provide frequency response and achieve normalized active power sharing. The column vector $\boldsymbol{\delta}(t)$ adjusts nodal frequency linearly according to power position $\boldsymbol{p}_p(t)$ as in (4), where $h$ is the control gain of droop-free controllers, and $\boldsymbol{p}_p(t)$ = $[p_{p1}(t) \; p_{p2}(t) \; ... \; p_{pN}(t)]^T$ is calculated in (5) to compare the normalized loading level of each DER to its neighbors. In (5), $\mathcal{N} = \{1, 2, \ldots, N\}$ is the set of nodes in the electrical network; $p_{ci}(t)$ and $p_{ci}^{ava}$ are respectively the active power output and the available capacity of controllable DER at node $i$; $W_{(i,j)}$ is the communication weight between nodes $i$ and $j$. For symmetric communication networks studied in this paper, $W_{(i,j)} = W_{(j,i)} > 0$ if there exists a direct communication link between nodes $i$ and $j$, and $W_{(i,j)} = W_{(j,i)} = 0$ otherwise.

$$\boldsymbol{\omega}(t) = \omega_0 \cdot \mathbf{1}_N + \boldsymbol{\delta}(t) \qquad (3)$$

$$\dot{\boldsymbol{\delta}}(t) = -h \cdot \boldsymbol{p}_c(t) \qquad (4)$$

$$p_{pi}(t) = \sum_{j \in \mathcal{N}, j \neq i} W_{(i,j)} \cdot \left( p_{ci}(t)/p_{ci}^{ava} - p_{cj}(t)/p_{cj}^{ava} \right) \qquad (5)$$

The neighboring communication process in (5) can be presented in a compact form (6). $\boldsymbol{L}_W$ is the $N$-dimensional Laplacian matrix of the weighted symmetric communication network; $\bar{\boldsymbol{p}}_c(t) = \boldsymbol{D}_p \cdot \boldsymbol{p}_c(t)$ contains the normalized active power states of the $N$ nodes, where $\boldsymbol{D}_p$ is an $N$-dimensional diagonal matrix defined as in (7) and $\boldsymbol{p}_c(t)$ is the vector of nodal active power injections from controllable DERs. With this, the O-NAPC based droop-free controller described in (3)-(7) can be presented in a compact form (8).

$$\boldsymbol{p}_p(t) = \boldsymbol{L}_W \cdot \bar{\boldsymbol{p}}_c(t) \qquad (6)$$

$$\boldsymbol{D}_p = diag\left(1/p_{c1}^{ava}, 1/p_{c2}^{ava}, \cdots, 1/p_{cN}^{ava}\right) \qquad (7)$$

$$\boldsymbol{\omega}(t) = \omega_0 \cdot \mathbf{1}_N - h \cdot \boldsymbol{L}_W \cdot \boldsymbol{D}_p \cdot \boldsymbol{p}_c(t) \qquad (8)$$

Finally, the state-space model of the O-NAPC based droop-free control is given in (9) by combining (1), (2), and (8).

$$\dot{\boldsymbol{p}}_{net}(t) = \boldsymbol{B} \cdot \left[ \boldsymbol{\omega}(t) - \omega_0 \cdot \mathbf{1}_N \right] = -h \cdot \boldsymbol{B} \cdot \boldsymbol{L}_W \cdot \boldsymbol{D}_p \cdot \boldsymbol{p}_c(t) \qquad (9)$$

Substituting $\boldsymbol{p}_c(t) = \boldsymbol{p}_{net}(t) - \boldsymbol{p}_u(t)$, the state-space model is rewritten as in (10). $\boldsymbol{p}_{net}(t)$ is the state vector denoting nodal net active power injections; $\boldsymbol{p}_u(t)$ denotes the vector of nodal active power injections from uncontrollable resources; $\boldsymbol{A}_{\text{O-NAPC}} = -h \cdot \boldsymbol{B} \cdot \boldsymbol{L}_W \cdot \boldsymbol{D}_p$ is the system matrix of O-NAPC. The O-NAPC based droop-free control loop is intuitively shown in Fig. 1 based on the derivation process (1)-(10).

$$\dot{\boldsymbol{p}}_{net}(t) = -h \cdot \boldsymbol{B} \cdot \boldsymbol{L}_W \cdot \boldsymbol{D}_p \cdot \left[ \boldsymbol{p}_{net}(t) - \boldsymbol{p}_u(t) \right] \qquad (10)$$

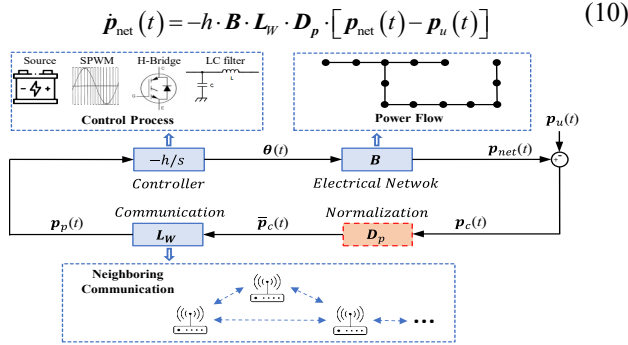

Fig. 1 Block diagram of the O-NAPC based droop-free control.

The O-NAPC based droop-free control (10) can achieve normalized active power sharing among controllable DERs. As the row sum of $\boldsymbol{L}_W$ is always zero, (11) is valid for any constant $\gamma$. It implies that $\dot{\boldsymbol{p}}_{net}(t)$ in (10) reaches zero (i.e., the steady state) when $\boldsymbol{D}_p \cdot [\boldsymbol{p}_{net}(\infty) - \boldsymbol{p}_u(\infty)] = \gamma \cdot \mathbf{1}_N$. Thus, the steady state of O-NAPC droop-free control can be described as in (12), indicating that active power outputs of the controllable DERs are unified at the same level in proportion to their available capacities, i.e., $p_{ci}(\infty)/p_{ci}^{ava} = p_{cj}(\infty)/p_{cj}^{ava}$.

$$\boldsymbol{L}_W \cdot \gamma \cdot \mathbf{1}_N = \mathbf{0}_N \qquad (11)$$

$$\boldsymbol{D}_p \cdot \left[ \boldsymbol{p}_{net}(\infty) - \boldsymbol{p}_u(\infty) \right] = \boldsymbol{D}_p \cdot \boldsymbol{p}_c(\infty) = \gamma \cdot \mathbf{1}_N \qquad (12)$$

## B. The A-NAPC Based Droop-Free Control

Different from the O-NAPC based droop-free control, an amplifier is further added in the droop-free control loop to form A-NAPC, as shown in Fig. 2. By applying the similar process that is adopted to derive the state-space model (10) of the O-NAPC based droop-free control in Section II.A, the state-space model of the A-NAPC based droop-free control can be presented as in (13), where $\boldsymbol{A}_{\text{A-NAPC}} = -h \cdot \boldsymbol{B} \cdot \boldsymbol{D}_p \cdot \boldsymbol{L}_W \cdot \boldsymbol{D}_p$ is the system matrix.

$$\dot{\boldsymbol{p}}_{net}(t) = -h \cdot \boldsymbol{B} \cdot \boldsymbol{D}_p \cdot \boldsymbol{L}_W \cdot \boldsymbol{D}_p \cdot \left[ \boldsymbol{p}_{net}(t) - \boldsymbol{p}_u(t) \right] \qquad (13)$$

Indeed, A-NAPC can achieve the same steady state as O-NAPC. Specifically, same as the steady-state operating point of O-NAPC, $\boldsymbol{D}_p \cdot [\boldsymbol{p}_{net}(\infty) - \boldsymbol{p}_u(\infty)] = \gamma \cdot \mathbf{1}_N$, enables $\boldsymbol{L}_W \cdot \boldsymbol{D}_p \cdot [\boldsymbol{p}_{net}(\infty) - \boldsymbol{p}_u(\infty)] = \mathbf{0}_N$, which will drive $\dot{\boldsymbol{p}}_{net}(\infty)$ in (13) also to zero. Thus, the steady states of A-NAPC and O-NAPC are identical.

On the other hand, as will be discussed in later sections and illustrated in case studies, O-NAPC and A-NAPC offer different stability margins for systems of various physical characteristics. Specifically, O-NAPC offers a larger stability margin for microgrids with more sparse communications, while A-NAPC is more suitable for microgrids with controllable DERs of smaller available capacities (e.g., $\boldsymbol{D}_p \succ 1$ p.u.) for achieving faster convergence performance.

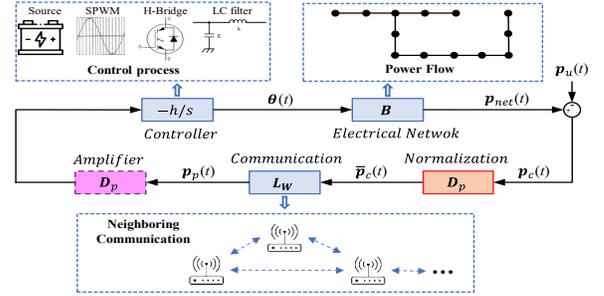

Fig. 2 Block diagram of the A-NAPC based droop-free control.

## III. Theoretical Stability Proof and Stability Margin Analysis of O-NAPC and A-NAPC Droop-Free Controls

This section first provides a thorough theoretical proof of the asymptotic stability of O-NAPC and A-NAPC based droop-free controls, followed by the quantitative evaluations of their stability margins to facilitate choosing proper control schemes according to microgrid physical specifications.

### A. Theoretical Proof of Asymptotic Stability

The asymptotic stability of the O-NAPC based droop-free control (10) and the A-NAPC based droop-free control (13) is thoroughly proved by testifying that all effective eigenvalues have negative real parts.

● *Eigenvalue Analysis for O-NAPC Based Droop-Free Control*: The following three lemmas are first introduced to support proving the asymptotic stability of the O-NAPC based droop-free control (10). Specifically, **Lemma 1** verifies $\text{Re}\{\lambda[(\boldsymbol{B}+\varepsilon\boldsymbol{I})\cdot(\boldsymbol{L}_W+\varepsilon\boldsymbol{I})\cdot\boldsymbol{D}_p]\} > 0$ by proving the real parts of eigenvalues for its inverse matrix are positive; **Lemma 2** validates $\text{Re}\{\lambda[(\boldsymbol{B}+\varepsilon\boldsymbol{I})\cdot(\boldsymbol{L}_W+\varepsilon\boldsymbol{I})\cdot\boldsymbol{D}_p\cdot(\boldsymbol{B}+\varepsilon\boldsymbol{I})\cdot(\boldsymbol{L}_W+\varepsilon\boldsymbol{I})\cdot\boldsymbol{D}_p]\} > 0$ by similar process of **Lemma 1**, which provides key support the proof of **Lemma 3**; **Lemma 3** proves that $\boldsymbol{B}\cdot\boldsymbol{L}_W\cdot\boldsymbol{D}_p$ has one simple zero eigenvalue and ($N$-1) eigenvalues with positive real parts by 2 steps: (i) Prove that matrix $\boldsymbol{B}\cdot\boldsymbol{L}_W\cdot\boldsymbol{D}_p$ has only one simple zero eigenvalue by showing there is only one trivial Jordan block with 1×1 dimension for $\lambda[\boldsymbol{B}\cdot\boldsymbol{L}_W\cdot\boldsymbol{D}_p]=0$; and (ii)

Prove that matrix $B \cdot L_w \cdot D_p$ has no other eigenvalue on the imaginary axis by contradiction, using the conclusion of as the opposite of the assumed proposition; ***Proposition*** 1 finally eliminates the zero eigenvalue, i.e. the asymptotic stability of the proposed droop-free control with O-NAPC is proved. As $B$ and $L_w$ are singular positive semi-definite (PSD) matrices, we introduce $\forall \varepsilon \in \mathbb{R}^+$ to augment them to positive definite (PD) matrices ($B+\varepsilon I$) and ($L_w+\varepsilon I$) and apply the continuity argument to assist the theoretical proof, where $I_N$ is an identity matrix of dimension $N$.

***Lemma*** 1: For $\forall \varepsilon \in \mathbb{R}^+$, $\text{Re}\{\lambda[(B+\varepsilon I)\cdot(L_w+\varepsilon I)\cdot D_p]\}>0$, where $D_p$ is a positive diagonal matrix, $\lambda[\cdot]$ are the eigenvalues of a matrix, and $\text{Re}\{\lambda[\cdot]\}$ are their real parts.

*Proof:* Since $\lambda[(B+\varepsilon I)\cdot(L_w+\varepsilon I)\cdot D_p]$ and $\lambda[D_p^{-1}\cdot(L_w+\varepsilon I)^{-1}\cdot(B+\varepsilon I)^{-1}]$ are reciprocal to each other as shown in (14), their real parts always have the same sign. Thus, we will study $\text{Re}\{\lambda[D_p^{-1}\cdot(L_w+\varepsilon I)^{-1}\cdot(B+\varepsilon I)^{-1}]\}$ instead.

$$\lambda\left[D_p^{-1}\cdot\left(L_w+\varepsilon I\right)^{-1}\cdot\left(B+\varepsilon I\right)^{-1}\right]=\lambda\left[\left\{(B+\varepsilon I)\cdot\left(L_w+\varepsilon I\right)\cdot D_p\right\}^{-1}\right] \quad (14)$$

For $0 \le \tau \le 1$, consider matrix $A(\tau)$ shown as follows:

$$A(\tau)=\left[(1-\tau)I+\tau D_p^{-1}\right]\cdot\left(L_w+\varepsilon I\right)^{-1}\cdot\left(B+\varepsilon I\right)^{-1} \quad (15)$$

We study the eigenvalues of $A(\tau)$ in two situations, i.e., $\tau=0$ and $0 < \tau \le 1$.

(*a*) When $\tau=0$, we have $A(0)=(L_w+\varepsilon I)^{-1}\cdot(B+\varepsilon I)^{-1}$. For $\forall \varepsilon > 0$, as $L_w+\varepsilon I$ and $B+\varepsilon I$ are PD matrices, $(L_w+\varepsilon I)^{-1}$ and $(B+\varepsilon I)^{-1}$ are also PD matrices. Thus, it is direct to conclude that $(L_w+\varepsilon I)^{-0.5}\cdot(B+\varepsilon I)^{-1}\cdot(L_w+\varepsilon I)^{-0.5}$ is a PD matrix. Moreover, $A(0)$ and $(L_w+\varepsilon I)^{-0.5}\cdot(B+\varepsilon I)^{-1}\cdot(L_w+\varepsilon I)^{-0.5}$ are similar matrices and thus have the same eigenvalues. Therefore, all the eigenvalues of $A(0)$ are positive, i.e., $\text{Re}\{\lambda[A(0)]\}>0$.

(*b*) For $\forall \tau \in (0,1]$, if $A(\tau)$ can be testified to have no eigenvalue on the imaginary axis, eigenvalues of $A(\tau)$ will stay positive, i.e., $\text{Re}\{\lambda[A(\tau)]\}>0$, for $\forall \tau \in (0,1]$. Therefore, it is direct to claim $\text{Re}\{\lambda[A(1)]\} = \text{Re}\{\lambda[D_p^{-1}\cdot(L_w+\varepsilon I)^{-1}\cdot(B+\varepsilon I)^{-1}]\} > 0$. Two scenarios are discussed as follows to prove that $A(\tau)$ for $\forall \tau \in (0,1]$ has no eigenvalue on the imaginary axis.

(i) Does $A(\tau)$ have zero eigenvalues? As both $(L_w+\varepsilon I)^{-1}$ and $(B+\varepsilon I)^{-1}$ are PD matrices, $\det[(L_w+\varepsilon I)^{-1}] > 0$ and $\det[(B+\varepsilon I)^{-1}] > 0$. With $\det[(1-\tau)I+\tau D_p^{-1}] > 0$, $\det[A(\tau)] > 0$ always holds. Thus, $A(\tau)$ will have no zero eigenvalues for $\forall \tau \in (0,1]$.

(ii) Does $A(\tau)$ have pure imaginary eigenvalues? Based on [22], the inverse of a nonsingular irreducible M-matrix is positive. Since $(L_w+\varepsilon I)$ and $(B+\varepsilon I)$ are nonsingular irreducible M-matrices, $(L_w+\varepsilon I)^{-1}$ and $(B+\varepsilon I)^{-1}$ are positive matrices. Therefore, it can be claimed that $\forall \tau \in (0,1]$, $A(\tau)$ is always a positive matrix. Thus, based on the Perron-Frobenius theorem [23, 24], we have $\lim_{k\to\infty}A^k(\tau)/r^k=v\cdot w^T$, where $r>0$, $v>0$, and $w>0$ are the Perron-Frobenius eigenvalue, the right eigenvector, and the left eigenvector of $A(\tau)$, respectively. $r$, $v$, and $w$ are all real-valued. Let $A_r(\tau) = A(\tau)/r$, we have $\lim_{k\to\infty}A^k(\tau)/r^k = \lim_{k\to\infty}A_r^k(\tau) = v\cdot w^T$. It indicates that the eigenvalues of $A_r^k(\tau)$ converge to (N-1) zeros and one positive real value that is equal to the summation of diagonal elements of $v\cdot w^T$.

Assume $\exists \tau_1 \in (0,1]$ such that $A(\tau_1)$ has a pair of pure imaginary eigenvalues $\pm i\cdot s$ with $s \in \mathbb{R}^+$. Then, $A_r(\tau_1)$ has a pair of pure imaginary eigenvalues $\pm i\cdot s_1 = \pm i\cdot(s/r)$. Hence, $A_r^2(\tau_1)$ will have a negative real eigenvalue $-s_1^2$. As $A_r^2(0)$ has the same eigenvalues with the PD matrix $(L_w+\varepsilon I)^{-0.5}\cdot(B+\varepsilon I)^{-1}\cdot(L_w+\varepsilon I)^{-1}\cdot(B+\varepsilon I)^{-1}\cdot(L_w+\varepsilon I)^{-0.5}/r^2$, it is direct to claim $\text{Re}\{\lambda[A_r^2(0)]\} > 0$. Thus, $\text{Re}\{\lambda[A_r^2(0)]\} > 0$ and $A_r^2(\tau_1)$ has a negative real eigenvalue $-s_1^2$. Based on the continuity of $A_r^2(\tau)$ w.r.t. $\tau$, $\exists \tau_2 \in (0, \tau_1)$ such that $A_r^2(\tau_2)$ has a pair of pure imaginary eigenvalues $\pm i\cdot s_2$ with $s_2 \in \mathbb{R}^+$. Then, $A_r^4(\tau_2) = [A_r^2(\tau_2)]^2$ will have a negative real eigenvalue $-s_2^2$.

Continue this process, for any $m \in \mathbb{Z}^+$, we can prove $\text{Re}\{\lambda[A_r^{2^m}(0)]\} > 0$ using the similar idea of proving $\text{Re}\{\lambda[A_r^2(0)]\} > 0$. Considering that $A_r^{2^m}(\tau_m)$ has a negative real eigenvalue $-s_m^2$ with $s_m \in \mathbb{R}^+$, there must $\exists \tau_{m+1} \in (0, \tau_m)$ such that $A_r^{2^m}(\tau_{m+1})$ has a pair of pure imaginary eigenvalues $\pm i\cdot s_{m+1}$ with $s_{m+1} \in \mathbb{R}^+$. Then, $A_r^{2^{m+1}}(\tau_{m+1})$ will have a negative real eigenvalue $-s_{m+1}^2$. In other words, $\exists \tau \in (0,1]$ such that the eigenvalues of $\lim_{m\to\infty, m\in\mathbb{Z}^+} A_r^{2^m}(\tau)$ will contain either a pair of pure imaginary values or a negative real value.

This contradicts the former claim that the eigenvalues of $\lim_{k\to\infty} A_r^k(\tau)$ will converge to (N-1) zeros and one positive real value. Therefore, $A(\tau)$ will have no pure imaginary eigenvalues for $\forall \tau \in (0,1]$.

Summarizing the above discussions (*a*) and (*b*), $A(\tau)$ for $\forall \tau \in [0,1]$ must have no pure imaginary eigenvalues. Hence, for $\forall \varepsilon \in \mathbb{R}^+$, $\text{Re}\{\lambda[(B+\varepsilon I)\cdot(L_w+\varepsilon I)\cdot D_p]\} > 0$ is proved. **Q.E.D.**

***Lemma*** 2: For $\forall \varepsilon \in \mathbb{R}^+$ and any positive diagonal matrix $D_p$, $\text{Re}\{\lambda[(B+\varepsilon I)\cdot(L_w+\varepsilon I)\cdot D_p\cdot(B+\varepsilon I)\cdot(L_w+\varepsilon I)\cdot D_p]\}>0$.

*Proof:* Let $A = (B+\varepsilon I)\cdot(L_w+\varepsilon I)\cdot D_p\cdot(B+\varepsilon I)\cdot(L_w+\varepsilon I)\cdot D_p$. Since $\lambda[A]$ and $\lambda[A^{-1}]$ are reciprocal to each other, their real parts always have the same sign and we will study $A^{-1}$ instead. That is, if $\text{Re}\{\lambda[A^{-1}]\} > 0$, we must have $\text{Re}\{\lambda[A]\} > 0$.

Let matrix $A_2(\tau) = \{[(1-\tau)I+\tau D_p^{-1}]\cdot(L_w+\varepsilon I)^{-1}\cdot(B+\varepsilon I)^{-1}\}^2$ for $\forall \tau \in [0,1]$. Notice that $A_2(\tau)=[A^{-1}(\tau)]^2$, thus ***Lemma*** 2 can be proved via the same process as ***Lemma*** 1. **Q.E.D.**

***Lemma*** 3: For any positive diagonal matrix $D_p$, $B \cdot L_w \cdot D_p$ has one simple zero eigenvalue and (N-1) eigenvalues with positive real parts.

*Proof:* We first prove that $B \cdot L_w \cdot D_p$ has one simple zero eigenvalue via the following two steps. The number of zero eigenvalues, i.e., the algebraic multiplicity of $\lambda[B\cdot L_w\cdot D_p]=0$, equals the sum of the sizes of Jordan blocks associated with $\lambda[B\cdot L_w\cdot D_p]=0$ in the Jordan normal form of $B\cdot L_w\cdot D_p$ [25].

(*a*) We show that there is only one trivial Jordan block for $\lambda[B\cdot L_w\cdot D_p]=0$. The null space of $B\cdot L_w\cdot D_p$ is given by the solutions to $B\cdot L_w\cdot D_p\cdot z=0_N$, $\forall z\in\mathbb{R}^N$. It has two possible solutions: $D_p\cdot z=1_N$ (i.e., $z=D_p^{-1}\cdot 1_N$) and $L_w\cdot D_p\cdot z=1_N$, due to the fact that $B$ and $L_w$ are both irreducible Laplacian matrices with null space $span\{1_N\}$. However, the second solution leads to a contradiction as $1_N^T\cdot L_w\cdot D_p\cdot z=1_N^T\cdot 1_N$ which implicates $0 = N$. Therefore, the eigenspace for $\lambda[B\cdot L_w\cdot D_p]$ is $span\{D_p^{-1}\cdot 1_N\}$, which is of one dimension. This also means that there is only one Jordan block associated with $\lambda[B\cdot L_w\cdot D_p]=0$.

(*b*) We further show that this Jordan block has dimension 1×1. Assuming the dimension of this Jordan block is $k\times k$ with $k>1$ instead, then there must exist a vector $u\ne 0_N$ such that $(B\cdot L_w\cdot D_p)^k\cdot u=0_N$, and $(B\cdot L_w\cdot D_p)^{k-1}\cdot u$ is an ordinary eigenvector of $B\cdot L_w\cdot D_p$ [25]. This indicates $(B\cdot L_w\cdot D_p)^{k-1}\cdot u = e\cdot D_p^{-1}\cdot 1_N$ for $e\in\mathbb{R}\backslash\{0\}$. By left-multiplying $1_N^T$ to both sides of $(B\cdot L_w\cdot D_p)^{k-1}\cdot u = e\cdot D_p^{-1}\cdot 1_N$, we can get the contradiction of $0 = 1_N^T\cdot D_p^{-1}\cdot 1_N$. Thus, $k=1$ must hold, indicating a trivial Jordan block.

By combining the above steps (*a*) and (*b*), we have proved that there is only one Jordan block of size 1×1 associated with $\lambda[B\cdot L_w\cdot D_p]=0$. Therefore, its algebraic multiplicity is one, i.e., $B \cdot L_W \cdot D_p$ has only one simple zero eigenvalue.

Next, we prove that all the other $(N-1)$ eigenvalues of $B \cdot L_W \cdot D_p$ have positive real parts. Based on **Lemma 1**, $\text{Re}\{\lambda[(B+\varepsilon I)\cdot(L_W+\varepsilon I)\cdot D_p]\} > 0$ holds for $\forall \varepsilon \in \mathbb{R}^+$. Thus, based on continuity of $\lambda[(B+\varepsilon I)\cdot(L_W+\varepsilon I)\cdot D_p]$ w.r.t. $\varepsilon$, when $\varepsilon = 0$, we must have $\text{Re}\{\lambda[B \cdot L_W \cdot D_p]\} \geq 0$.

For $\text{Re}\{\lambda[B \cdot L_W \cdot D_p]\} = 0$, there exist two situations:

(*a*) Does $B \cdot L_W \cdot D_p$ have zero eigenvalues? Based on the above discussions, $B \cdot L_W \cdot D_p$ has only one zero eigenvalue.

(*b*) Does $B \cdot L_W \cdot D_p$ have pure imaginary eigenvalues? Assume $\exists s \neq 0$ such that $(i \cdot s)$ is an eigenvalue of $B \cdot L_W \cdot D_p$, then $B \cdot L_W \cdot D_p \cdot B \cdot L_W \cdot D_p$ will have a negative real eigenvalue $(i \cdot s)^2 = -s^2 < 0$. However, based on **Lemma 2**, $\forall \varepsilon \in \mathbb{R}^+$, $\text{Re}\{\lambda[(B+\varepsilon I)\cdot(L_W+\varepsilon I)\cdot D_p \cdot (B+\varepsilon I)\cdot(L_W+\varepsilon I)\cdot D_p]\} > 0$. Thus, based on continuity w.r.t. $\varepsilon$, when $\varepsilon = 0$, we have $\text{Re}\{\lambda[B \cdot L_W \cdot D_p \cdot B \cdot L_W \cdot D_p]\} \geq 0$. The contradiction implies the assumed eigenvalue $(i \cdot s)$ does not exist, and $B \cdot L_W \cdot D_p$ must not have pure imaginary eigenvalues.

Thus, the above steps (*a*) and (*b*) together indicate that $\lambda[B \cdot L_W \cdot D_p]$ is either zero or has a positive real part.

In summary, by combining the above two-part proof, it is direct to claim that $B \cdot L_W \cdot D_p$ has one simple zero eigenvalue and $(N-1)$ eigenvalues with positive real parts. **Q.E.D.**

With **Lemma 3**, the asymptotic stability of the O-NAPC based droop-free control (10) can be readily verified via **Proposition 1**.

**Proposition 1**: All the $(N-1)$ effective eigenvalues of the system matrix $A_{\text{O-NAPC}} = -h \cdot B \cdot L_W \cdot D_p$ have negative real parts, and the O-NAPC based droop-free controlled microgrid (10) is asymptotically stable.

*Proof*: **Lemma 3** indicates that $B \cdot L_W \cdot D_p$ has one simple zero eigenvalue and $(N-1)$ eigenvalues with positive real parts. Thus, by multiplying the negative gain $-h$, the system matrix $A_{\text{O-NAPC}} = -h \cdot B \cdot L_W \cdot D_p$ has $N$ eigenvalues with non-positive real parts, among which only one eigenvalue is zero. The power balance and dynamic balance constraints (16) always hold in the microgrids, according to equations (1) and (9). Indeed, equation (16) describes the linear dependence between variables, i.e., $p_{\text{net}}^i(t)$ and $\dot{p}_{\text{net}}^i(t)$ of node $i$ can be equivalently substituted by a linear combination of variables from all other nodes without affecting system dynamics and the other $(N-1)$ non-zero eigenvalues. With this, the system dimension can be reduced to $(N-1)$ to eliminate the only zero eigenvalue.

$$\sum_{i \in N} p_{\text{net}}^i(t) = 0, \sum_{i \in N} \dot{p}_{\text{net}}^i(t) = 0 \qquad (16)$$

Therefore, the system matrix $A_{\text{O-NAPC}}$ has $(N-1)$ effective eigenvalues, all with negative real parts. Thus, the O-NAPC based droop-free control is asymptotically stable. **Q.E.D.**

• *Eigenvalue Analysis for A-NAPC Based Droop-Free Control*: The following two lemmas are discussed to facilitate proving the asymptotic stability of the A-NAPC based droop-free control (13). Specifically, **Lemma 4** proves that all eigenvalues of the system matrix $A_{\text{A-NAPC}} = -h \cdot B \cdot D_p \cdot L_W \cdot D_p$ are non-positive real values. **Lemma 5** proves that the system matrix $A_{\text{A-NAPC}} = -h \cdot B \cdot D_p \cdot L_W \cdot D_p$ has one simple zero eigenvalue. And **Proposition 2** eliminates the zero eigenvalue and completes the proof for asymptotic stability of the droop-free control with A-NAPC.

**Lemma 4**: All eigenvalues of the system matrix $A_{\text{A-NAPC}} = -h \cdot B \cdot D_p \cdot L_W \cdot D_p$ are non-positive real values.

*Proof*: As matrix $B$ is a PSD matrix, it can be represented as $B = C_B \cdot C_B^T$ via Cholesky decomposition. Based on the fact that when several square matrices are multiplied, the eigenvalues of the final product remain the same when switching the orders of these square matrices cyclically, we have $\lambda[B \cdot D_p \cdot L_W \cdot D_p] = \lambda[C_B \cdot C_B^T \cdot D_p \cdot L_W \cdot D_p] = \lambda[C_B^T \cdot D_p \cdot L_W \cdot D_p \cdot C_B]$.

Let $x = D_p \cdot C_B \cdot x$ for $x \neq 0_N$, we have

$$x^T \cdot C_B^T \cdot D_p \cdot L_W \cdot D_p \cdot C_B \cdot x = (D_p \cdot C_B \cdot x)^T \cdot L_W \cdot (D_p \cdot C_B \cdot x) \qquad (17)$$
$$= y^T \cdot L_W \cdot y \geq 0$$

As $L_W$ is a symmetric PSD matrix, (17) implies that $C_B^T \cdot D_p \cdot L_W \cdot D_p \cdot C_B$ is also a symmetric PSD, i.e., the eigenvalues of $B \cdot D_p \cdot L_W \cdot D_p$ are all non-negative. Hence, with the negative control gain $-h$, all eigenvalues of the system matrix $A_{\text{A-NAPC}}$ are non-positive real values. **Q.E.D.**

**Lemma 5**: The system matrix $A_{\text{A-NAPC}} = -h \cdot B \cdot D_p \cdot L_W \cdot D_p$ has one simple zero eigenvalue.

*Proof*: This lemma can be proved using the same idea as the first part of the proof for **Lemma 3**: (i) We can prove that there is only one trivial Jordan block for $\lambda[B \cdot D_p \cdot L_W \cdot D_p]=0$; and (ii) We can further prove that this Jordan block has dimension 1×1. These two steps together indicate that there is only one Jordan block of size 1×1 associated with $\lambda[B \cdot D_p \cdot L_W \cdot D_p]=0$. Therefore, its algebraic multiplicity is one, i.e., $B \cdot D_p \cdot L_W \cdot D_p$ has only one simple zero eigenvalue. **Q.E.D.**

With **Lemmas 4** and **5**, the asymptotic stability of the A-NAPC based droop-free control (13) can be readily verified via **Proposition 2**.

**Proposition 2**: All the $(N-1)$ effective eigenvalues of the system matrix $A_{\text{A-NAPC}} = -h \cdot B \cdot D_p \cdot L_W \cdot D_p$ are negative real values, and the A-NAPC based droop-free controlled microgrid (13) is asymptotically stable.

*Proof*: **Lemma 4** and **Lemma 5** collectively indicate that all $N$ eigenvalues of the system matrix $A_{\text{A-NAPC}}$ are non-positive real values, among which only one eigenvalue is zero. Same as **Proposition 1**, the power balance and dynamic balance constraints (16) always hold and the system dimension can be reduced to $(N-1)$ to eliminate the only zero eigenvalue.

Therefore, the system matrix $A_{\text{A-NAPC}}$ has $(N-1)$ negative real-valued eigenvalues, and the A-NAPC based droop-free control is asymptotically stable. **Q.E.D.**

### B. Stability Margin Analysis

As both O-NAPC and A-NAPC based droop-free controls are asymptotically stable according to **Propositions 1** and **2**, their stability margins are further analyzed to facilitate choosing proper control schemes based on the system's physical characteristics for achieving faster convergence performance. The stability margin and rate of convergence are closely related to the system's dominant pole (i.e., the non-zero eigenvalue closest to the imaginary axis). That is, the control scheme whose real part of the dominant pole has a larger absolute value presents a higher stability margin and thus a faster rate of convergence. However, the system specifications can be easily altered by faults and cyber attacks, which directly affect the stability margin. Thus, the vulnerability of NAPC based droop-free control is further analyzed to identify the weakness in the microgrids.

• *The Influence Factors of Stability Margin*

The state-space models (10) and (13) involve three matrices that describe system specifications: the susceptance matrix of electrical network $B$, the Laplacian matrix of communication network $L_W$, and the diagonal matrix $D_p$ describing available capacities of controllable DERs. Three metrics are defined in this paper to quantify their impacts on the dominant pole and thus the stability margin of the system.

(i) *The average impedance in matrix $\boldsymbol{B}$*: The average impedance $B_{avg}$ is defined as in (18), where $N_l$ is the number of electric lines in the microgrid. Note that $N_l$ is equal to ($N$-1) for radial electrical networks. As impedance represents the electrical distance between nodes in the system and is usually proportional to the actual lengths of feeders, the average impedance metric could reflect the average actual distance between nodes in the system.

$$B_{\text{avg}} = tr(\boldsymbol{B}) / 2N_l \qquad (18)$$

(ii) *The sparseness of matrix $\boldsymbol{L_W}$*: The sparseness metric of matrix $\boldsymbol{L_W}$ is defined as in (19). The total number of edges in a complete graph with $N$ nodes is $N \cdot (N-1)/2$, and the number of edges in the communication network is $tr(\boldsymbol{L_W})/2$. Thus, the metric $L_{\text{sparse}}$ quantifies the relative difference in the number of edges between the communication network and the complete graph, describing the sparseness of the communication network. Generally, a denser communication network can stabilize the system faster after disturbances, as more information is exchanged among DERs.

$$L_{\text{sparse}} = 1 - tr(\boldsymbol{L_W}) / \left[ N(N-1) \right] \qquad (19)$$

(iii) *The average available capacity of controllable DERs*: The average available capacity $P_{avg}$ of controllable DERs is calculated as in (20). Indeed, the available capacity of controllable DERs is a critical factor for system stability and convergence performance. Specifically, DERs with larger available capacities would normally offer a higher system stability margin.

$$P_{\text{avg}} = tr(\boldsymbol{D}_p^{-1}) / N \qquad (20)$$

In real microgrids, the dominant pole is actually affected by these three metrics collectively, indicating that the influences of these three metrics on the stability margins of the two droop-free control schemes are rather complicated. Thus, the stability margins can be vulnerable to the perturbation of system parameters.

- *The Vulnerability Analysis*

In practice, the operation of microgrids experiences random disturbances that could severely impact the dynamic performance or even lead to system instability. The vulnerability analysis can identify the weakness of the system, i.e., the components that are most susceptible to causing system instability following a disturbance. This can guide operation and maintenance staff to enhance monitoring and strengthen the infrastructure of the weak components.

The perturbations of three matrices involved in the system matrix, i.e., $\boldsymbol{B}$, $\boldsymbol{L_W}$, and $\boldsymbol{D_p}$, are utilized in this subsection to represent disturbances in microgrid operations. The impedance will slowly increase due to the effect of the electrical line aging. The communication weights could be perturbed by operational errors and cyber-physical attacks [26]. It can also represent communication errors in the neighboring information exchange, i.e., the impact of a 10% communication error in the neighboring information exchange can be considered as a 10% communication weights deviation. The capacities of controllable DERs will change due to faults or end-user behavior. Electrical faults may lead to the capacities of controllable DERs being partially/fully unavailable, and some end-users could be randomly disconnected from microgrids.

The system vulnerability with respect to perturbations of electrical line susceptance, communication weights, and capacities of controllable DERs is further analyzed in this subsection.

The vulnerability analysis is to identify the weakness in the system, i.e., locating the minimum perturbations of matrices $\boldsymbol{B}$, $\boldsymbol{L_W}$, and $\boldsymbol{D_p}$ that can change the dominant pole to non-negative and destabilize the system. The perturbed dominant pole is calculated as in (21).

$$\lambda_p = \lambda_d + \partial \lambda_d / \partial \sigma_{i,j} \cdot \rho \qquad (21)$$

where $\lambda_p$ is the perturbed dominant pole, $\lambda_d$ is the original dominant pole, $\partial \lambda_d / \partial \sigma_{i,j}$ is the sensitivity of $\lambda_d$ w.r.t. the perturbation on the $(i, j)$-th element of the perturbed matrix, and $\rho$ describes the amount of the perturbation.

From (21), it is clear that the minimum perturbation $\sigma$ that can destabilize the system is achieved with the maximum eigenvalue sensitivity $\partial \lambda_d / \partial \sigma_{i,j}$. The eigenvalue sensitivity can be calculated as in (22) [27].

$$\partial \lambda_d / \partial \sigma_{i,j} = \left[ \boldsymbol{\psi}_d^T \left( \partial A / \partial \sigma_{i,j} \right) \boldsymbol{\phi}_d \right] / \left( \boldsymbol{\psi}_d^T \cdot \boldsymbol{\phi}_d \right) \qquad (22)$$

where $\boldsymbol{\psi}_d$ and $\boldsymbol{\phi}_d$ are the left and right eigenvectors of $\lambda_d$, respectively. Furthermore, the perturbed system matrix $A$ is an explicit function of $\sigma_{i,j}$. For example, if the communication weight of the O-NAPC control scheme is perturbed, the perturbed system matrix $A_{\text{O-NAPC}\_p}$ is shown as.

$$A_{\text{O-NAPC}\_p} = -h \cdot \boldsymbol{B} \cdot (\boldsymbol{L_W} + \boldsymbol{\sigma}) \cdot \boldsymbol{D}_p \qquad (23)$$

where $\boldsymbol{\sigma}$ is the perturbation matrix. The $(i, j)$-th element of $\boldsymbol{\sigma}$ is $\sigma_{i,j}$, while the other elements are all zero. The partial derivative of $A_{\text{O-NAPC}\_p}$ w.r.t. $\sigma_{i,j}$ can be obtained as in (24).

$$\begin{aligned} \partial A_{\text{O-NAPC}\_p} / \partial \sigma_{i,j} &= -h \cdot \partial \left[ \boldsymbol{B} \cdot (\boldsymbol{L_W} + \boldsymbol{\sigma}) \cdot \boldsymbol{D}_p \right] / \partial \sigma_{i,j} \\ &= -h \cdot \partial \left[ \boldsymbol{B} \cdot \boldsymbol{L_W} \cdot \boldsymbol{D}_p \right] / \partial \sigma_{i,j} - h \cdot \partial \left[ \boldsymbol{B} \cdot \boldsymbol{\sigma} \cdot \boldsymbol{D}_p \right] / \partial \sigma_{i,j} \\ &= -h \cdot \partial \left[ \boldsymbol{B} \cdot \boldsymbol{\sigma} \cdot \boldsymbol{D}_p \right] / \partial \sigma_{i,j} \end{aligned} \qquad (24)$$

From (21)-(24), the eigenvalue sensitivity with respect to perturbation of the $(i, j)$-th element of matrices $\boldsymbol{B}$, $\boldsymbol{L_W}$, and $\boldsymbol{D_p}$ can be directly calculated. Based on the eigenvalue sensitivity, the weakness in the microgrids can be identified. The components with the maximum eigenvalue sensitivity are the vulnerable points in the system, which could provide valuable guidelines for enhancing stability or defending against external attacks. The strengthening approaches could include applying more frequent maintenance and patrol for the weak component, using more advanced encrypted communication technology, and updating the infrastructure of electrical and communication networks.

## IV. CASE STUDY

This section conducts multiple case studies on a HIL microgrid testbed to validate the asymptotic stability of the two NAPC-based droop-free control schemes as well as the effectiveness of the proposed vulnerability analysis. The entire HIL microgrid testbed is shown in Fig. 3, which consists of a physical part of the electrical system, a digital part of the electrical system, the control system, and the console of the microgrid. The physical part of the electrical system includes a Ranogy battery storage, an Imperix PEH2015 H-bridge, a designed LC filter, and a steamer load. The digital part of the electrical system and the control system are deployed on OP 5707 simulator, which contains an Intel Xeon E5 CPU and a Virtex-7 field programmable gate array (FPGA) board. The control system calculation is conducted on the CPU at a time step of $20\mu s$. The digital part of the electrical system, including line impedance, PV panels, DC/DC, and converters to enrich the microgrid setup, is simulated on the FPGA at a time step of $0.5\mu s$ [14]. The console is illustrated on MATLAB Simulink, which is convenient and clear for dynamic performance monitoring and demonstration.

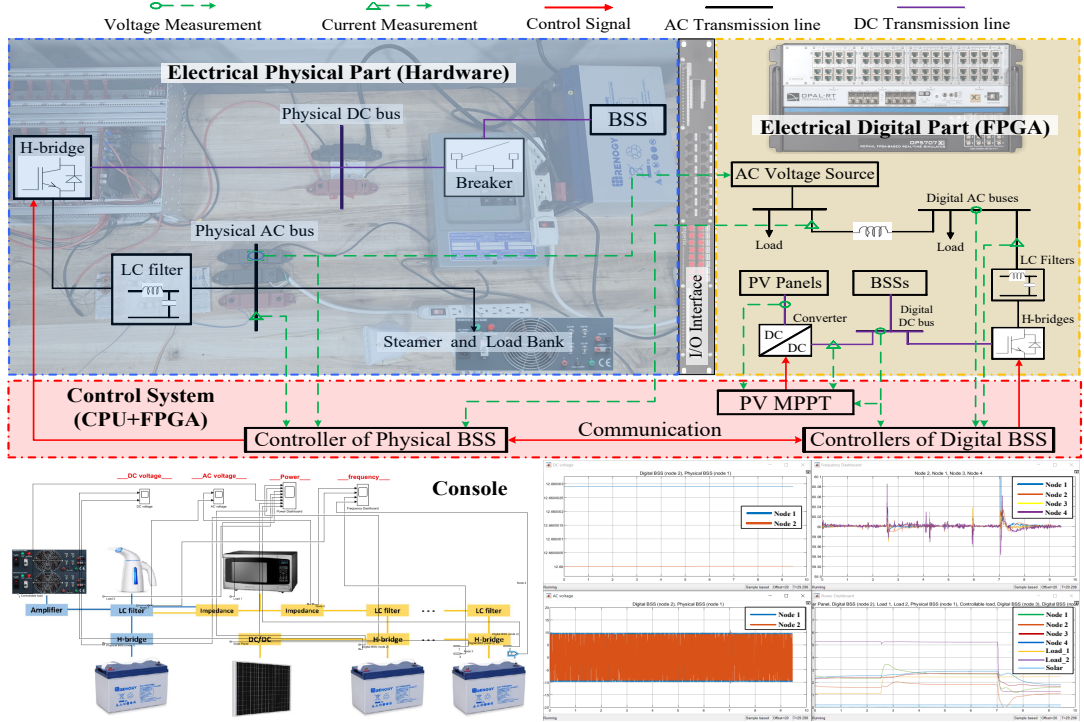

Fig. 3 Integrated HIL microgrid testbed.

The electrical physical part, electrical digital part, and control system are integrated together by OP5707 simulator, forming a high-performance HIL microgrid testbed. (i) Physical part and control system integration: the current and voltage measurements on the electrical physical part are fed into the control system calculation on CPU via analog signals, while the control system generates digital control signals to the converter interface to control the physical part; (ii) Physical part and digital part integration: the nodal voltage states of the electrical physical part are sent to FPGA via analog signals, which are assigned to the AC voltage-source in the digital part. With this, the voltage dynamics of the physical part is modeled in the digital part to deliver its impact on the digital electrical network, and the current state information of the digital part is further leveraged to adjust the control of the physical part.

First, the dynamics and steady-state performances of O-NAPC and A-NAPC are compared with APC-based droop-free control via two cases. In each case, system parameters (i.e., $\boldsymbol{B}$, $\boldsymbol{L}_W$, and $\boldsymbol{D}_p$) and control gain $h$ of the three control schemes are set the same, while different dominant poles show the influences of distinct control schemes on system dynamics. The convergence condition is given in (25) with predefined parameters $\Delta t$ and $\xi$, i.e., the minimum $t_c$ satisfying (25) is identified as the convergence time after disturbances.

$$\min \ \left\{ t \big\| \ \boldsymbol{p}_{\text{net}}\left(t\right) - \boldsymbol{p}_{\text{net}}\left(t - \Delta t\right) \right| \leq \xi \right\} \quad (25)$$

Second, stability margin analysis is further demonstrated in terms of the proposed influence factors and the vulnerability against perturbations. Influences of the three proposed metrics (i.e., $B_{avg}$, $L_{sparse}$, and $P_{avg}$) on the stability margin are illustrated to show different scenarios suitable for O-NAPC and A-NAPC schemes. The effectiveness of the proposed vulnerability analysis is validated by calculating the eigenvalue sensitivity to every non-zero element in the matrices $\boldsymbol{B}$, $\boldsymbol{L}_W$, and $\boldsymbol{D}_p$, showing the weakness of O-NAPC and A-NAPC based droop-free control schemes against faults or cyber-physical attacks.

### A. Control Performance Comparison of O-NAPC, A-NAPC, and APC-Based Droop-free Control Schemes

Two cases are studied in this subsection to demonstrate the convergence performance and consensus equilibrium of the three control schemes, as shown in Fig. 4, where $p_i^c$ is the active power of the $i$-th controllable DER, and $p_i^u$ is the uncontrollable active power of the $i$-th node, i.e., load, random active power disturbances, and uncontrollable DERs, etc. The microgrids studied in both cases have 4 nodes, but with different matrices $\boldsymbol{B}$, $\boldsymbol{L}_W$, and $\boldsymbol{D}_p$ to illustrate distinct dynamic characteristics of the three control schemes in microgrids with different system specifications.

The 4-node test system in Case 1 shown in Fig. 4(*a*) has a cycle communication network. Each node has one controllable DER, and their capacities are listed in TABLE I. The base value is 100kW. Other essential parameters of the control system are given in Table II. The test system for Case 2 is shown in Fig. 4(*b*). The electrical network topology of Case 2 is the same as Case 1, but with different line impedances. The communication network of Case 2 is a complete graph, implying that each node communicates with all other three nodes. To show the dynamic trajectory to approaching the (normalized) active power consensus, in both cases, random load injection into Node 4 at 5s and random solar power injection into Node 3 at 15s are considered as representative uncontrollable disturbances. The disturbances that will change the topology of electrical networks or communication networks are not discussed in this paper, as electrical networks and communication networks need to be connected graphs. Handling line disconnections will be investigated in our future work.

TABLE I THE AVAILABLE CAPACITIES OF DERS IN THE TWO TEST CASES

| Node | 1 | 2 | 3 | 4 |
|---|---|---|---|---|
| Case 1 (p.u.) | 1.85 | 1.73 | 1.15 | 1.67 |
| Case 2 (p.u.) | 7.67 | 1.58 | 1.52 | 1.11 |

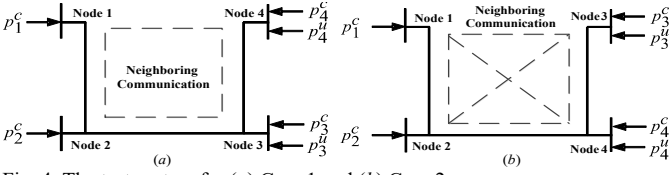

Fig. 4  The test system for (*a*) Case 1 and (*b*) Case 2.

TABLE II  OTHER ESSENTIAL PARAMETERS OF THE CONTROL SYSTEM

| Component | Parameters | Value |
|---|---|---|
| PQ calculator filter | Time constant (ms) | 1.00 |
| SPWM | Frequency (Hz) | 50000 |
| LC filter | Inductance (mH) | 4.00 |
|  | Capacitance (μF) | 141 |
|  | Resistance (Ω) | 1.72 |
| Line | Inductance (mH) | 5.30 |
| Inner current controller | Proportional gain | 0.15 |
|  | Integral gain | 6.6 |
| Outer voltage controller | Proportional gain | 12 |
|  | Integral gain | 200 |
| Droop-free controller | Control gain | 1.20 |

The dynamics of O-NAPC, A-NAPC, and APC for Case 1 are shown in Fig. 5-Fig. 7, respectively. As shown in Fig. 5 and Fig. 6, both O-NAPC and A-NAPC control schemes have the same active power output at 2.0p.u. before the load disturbance occurs at 5s. After the load disturbance, both control schemes achieve the same normalized active power consensus at around 2.227 p.u.. Specifically, the load sharing is 0.227×127kW=28.83kW for node 1, 0.227×119kW=27.01kW for node 2, 0.227×79kW=17.93kW for node 3, and 0.227×115kW=26.11kW for node 4. After the disturbance of solar power injection occurs at 15s, the normalized active power consensus becomes 1.477 p.u., and the active power reduction shares are 95.25kW, 89.25kW, 59.25kW, and 86.25kW for the four nodes. In other words, the amounts of load burden shared by individual nodes have the same percentage with respect to their available capacities. Different from O-NAPC and A-NAPC, the APC control scheme achieves a steady state by equally sharing the load burden among the controllable DERs. As shown in Fig. 7, the disturbance shares of individual nodes are approximately 25 kW after the load increase and 82.5 kW after the solar power injection. The frequency dynamics of O-NAPC, A-NAPC, and APC control schemes for Case 1 are shown in Fig. 8. The frequency of these three control schemes can be stabilized after disturbances, and they have similar dynamic characteristics with the average nodal frequency being around the system frequency of 60 Hz.

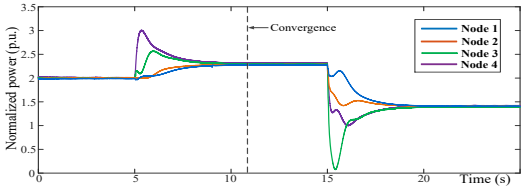

Fig. 5  Dynamics of O-NAPC for Case 1.

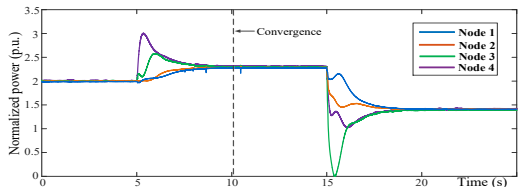

Fig. 6  Dynamics of A-NAPC for Case 1.

The dynamics of O-NAPC, A-NAPC, and APC for Case 2 are shown in Fig. 9-Fig. 11, respectively. Similar to Case 1, Fig. 9 and Fig. 10 show that both O-NAPC and A-NAPC achieve the same steady state of equal normalized active power. Fig. 11 further indicates that APC will operate controllable DERs at the same active power output level regardless of their available capacities. The frequency dynamics of O-NAPC, A-NAPC, and APC control schemes for Case 2 are shown in Fig. 12. The frequency of these three control schemes for Case 2 is similar to Case 1, with stabilized frequency after disturbances and average nodal frequency around 60Hz.

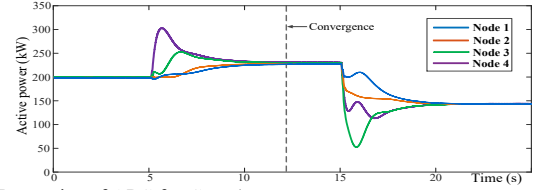

Fig. 7  Dynamics of APC for Case 1.

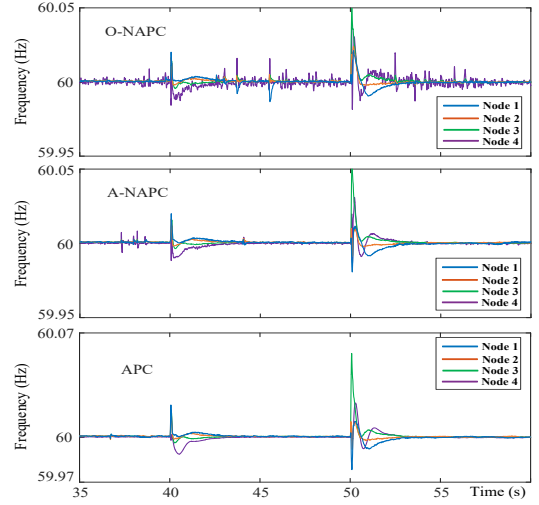

Fig. 8  Frequency dynamics for Case 1.

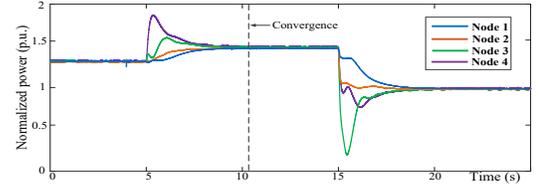

Fig. 9  Dynamics of O-NAPC for Case 2.

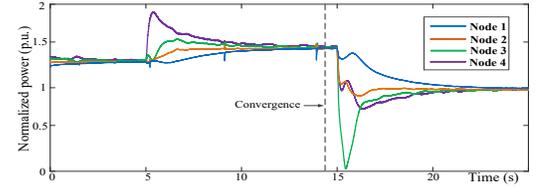

Fig. 10  Dynamics of A-NAPC for Case 2.

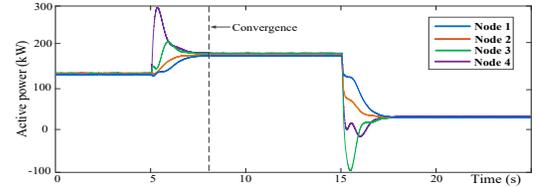

Fig. 11  Dynamics of APC for Case 2.

The real parts of the dominant poles and the convergence time of the three control schemes in the two cases are further compared in TABLE III. In Case 1, the convergence speeds follow APC < O-NAPC < A-NAPC; in comparison, the convergence speeds in Case 2 follow A-NAPC < O-NAPC < APC. A-NAPC converges the fastest in Case 1, because of its larger controller inputs $p_p(t)$. Since O-NAPC and A-NAPC include a normalization process in the control loop, with controllable DERs of smaller capacities, their $p_p(t)$ are amplified compared to APC. On the contrary, in Case 2, APC has the fastest convergence speed. This is because the available capacities of controllable DERs in Case 2 are much larger than those in Case 1, which reduces $p_p(t)$ of O-NAPC and A-NAPC.

Besides, the participation factors of O-NAPC, A-NAPC, and APC schemes are demonstrated in TABLE IV and TABLE V. The participation vector $P$ is given by $P = \psi_d^T \odot \phi_d$, where $\psi_d$ and $\phi_d$ are the left and right eigenvectors of dominant pole $\lambda_d$, respectively. The participation factors in TABLE IV and TABLE V collectively show that the state $p_{net,1}$ has the most significant influence on the system dominant pole. Although with O-NAPC and A-NAPC control schemes, node 1 with the largest capacity seems to have the dominant effect, the similar participation factors of the APC scheme show that system dominant poles are not solely driven by the available capacities of nodes. Indeed, it reflects a collective effect of the electrical network, the communication network, and the available power capacities of individual nodes.

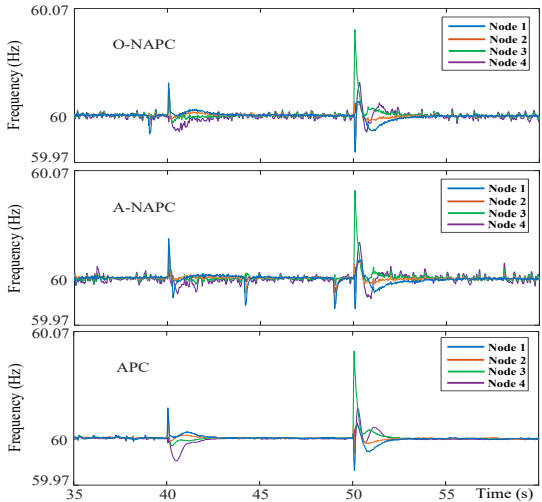

Fig. 12 Frequency dynamics for Case 2.

TABLE III  Comparison of Real Parts of Dominant Poles and Convergence Time

| Case | O-NAPC | | A-NAPC | | APC | |
|---|---|---|---|---|---|---|
| | Dominant Pole | Convergence Time ($s$) | Dominant Pole | Convergence Time ($s$) | Dominant Pole | Convergence Time ($s$) |
| 1 | -1.61 | 10.9 | -1.87 | 10.1 | -1.38 | 12.2 |
| 2 | -2.99 | 10.3 | -1.25 | 14.2 | -4.42 | 7.9 |

TABLE IV  Participation Factors of Droop-Free Control in Case 1

| States | $p_{net,1}$ | $p_{net,2}$ | $p_{net,3}$ | $p_{net,4}$ |
|---|---|---|---|---|
| O-NAPC | 0.7095 | 0.0666 | 0.1329 | 0.0910 |
| A-NAPC | 0.7101 | 0.0753 | 0.1187 | 0.0959 |
| APC | 0.7481 | 0.0389 | 0.1527 | 0.0603 |

TABLE V  Participation Factors of Droop-Free Control in Case 2

| States | $p_{net,1}$ | $p_{net,2}$ | $p_{net,3}$ | $p_{net,4}$ |
|---|---|---|---|---|
| O-NAPC | 0.3543 | 0.2309 | 0.2351 | 0.1797 |
| A-NAPC | 0.3546 | 0.2443 | 0.2405 | 0.1606 |
| APC | 0.7492 | 0.0706 | 0.0832 | 0.0970 |

In summary, the A-NAPC control scheme is more suitable for microgrids with smaller available capacities of controllable DERs. Due to the normalization process and extra amplifier, smaller available capacities of controllable DERs will increase the absolute value of dominant poles for the A-NAPC control scheme. In comparison, the APC control scheme presents a faster convergence speed for microgrids with controllable DERs of larger available capacities, which will increase the absolute value of dominant poles for the APC control scheme. It is worth noting that for the APC control scheme, active power outputs cannot exceed the available capacity of the smallest DER because the load burdens shared among controllable DERs are always identical. Thus, the application of the APC control scheme is limited since the system can potentially be unstable if the disturbances trigger the violation of the capacity limit of any controllable DER. With this, in practice, the NAPC control schemes are preferred for microgrids with heterogeneous controllable DERs.

### B. The Sensitivity Analysis of $B_{avg}$, $L_{sparse}$, and $P_{avg}$

As shown in TABLE III, the convergence speed is dominated by the system's dominant poles. For a stable microgrid with $(N-1)$ negative eigenvalues, the system's dominant poles can be used to approximate the convergence speed of system dynamics against disturbances. Thus, the absolute value of the real part of the system's dominant poles can be considered as the stability margin. The sensitivity of the three proposed metrics $B_{avg}$, $L_{sparse}$, and $P_{avg}$ on stability margin is further analyzed for systems of different sizes.

Take $B_{avg}$ as an example to show the process of the sensitivity analysis. First, a certain number of samples with different electrical network topologies and different line impedance are generated with the same value of $B_{avg}$; then, the same process is repeated for other values of $B_{avg}$. Note that when calculating the stability margin for $B_{avg}$, the matrices $L_W$ and $D_p$ remain unchanged. With this, the average stability margin w.r.t. $B_{avg}$ for different system sizes is obtained in a statistical manner. Similarly, the average stability margin w.r.t. $L_{sparse}$ and $P_{avg}$ can be obtained.

As shown in Fig. 13, the stability margin of the two NAPC schemes has the same changing trend w.r.t. $B_{avg}$ for systems of different sizes, where O represents O-NAPC control scheme and A refers to A-NAPC control scheme. A lower $B_{avg}$ value means that the electrical distance between nodes is farther, while a higher $B_{avg}$ value indicates that the electrical connection between nodes is stronger. Hence, decreasing the impedance of electrical lines will strengthen the system stability margin.

Similarly, Fig. 14 shows that as $L_{sparse}$ increases, the stability margin of both control schemes decreases and shares a similar trend. It implies that a dense communication network will increase the system stability margin due to more active power information exchanged between neighboring nodes.

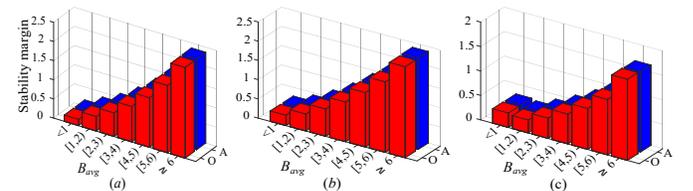

Fig. 13  The stability margin of the two control schemes w.r.t. $B_{avg}$: (a) The system size is 8, (b) the system size is 16, (c) The system size is 32.

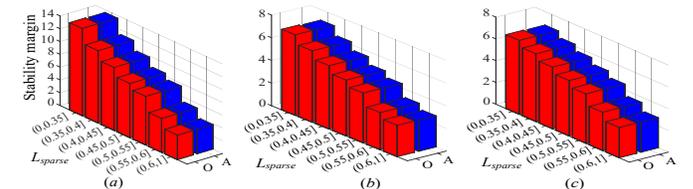

Fig. 14  The stability margin of the two control schemes w.r.t. $L_{sparse}$: (a) The system size is 8, (b) The system size is 16, (c) The system size is 32.

The stability margin of the two control schemes w.r.t. $P_{\text{avg}}$ is illustrated in Fig. 15. Different from $B_{\text{avg}}$ and $L_{\text{sparse}}$, the A-NAPC control scheme has a higher stability margin when $P_{\text{avg}}$ is smaller, and the O-NAPC control scheme has a higher stability margin when $P_{\text{avg}}$ is larger. It should be noted that as $P_{\text{avg}}$ increases for systems of different sizes, the system stability margin decreases for both NAPC-based control schemes. This is because the increased $P_{\text{avg}}$ leads to decreased $p_p(t)$ in (5), i.e., the active power information exchanged through neighboring communication is weakened. For maintaining the stability margin and accelerating the convergence after disturbances, the control gain $h$ needs to be increased as the system size and $P_{\text{avg}}$ increase.

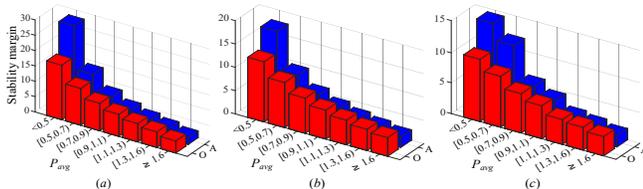

Fig. 15 The stability margin of the two control schemes w.r.t. $P_{\text{avg}}$: (*a*) The system size is 8, (*b*) The system size is 16, (*c*) The system size is 32.

### C. The Vulnerability Analysis

The vulnerability analysis is conducted on an 8-node microgrid for the two NAPC control schemes concerning perturbations of matrices $\boldsymbol{B}$, $\boldsymbol{L}_W$, and $\boldsymbol{D}_p$. The vulnerability analysis of the two NAPC control schemes on the perturbations of matrix $\boldsymbol{B}$ is shown in Fig. 16. The perturbations of matrix $\boldsymbol{B}$ reflect the change of parameters for electrical lines due to routine maintenance or component aging. Therefore, the $(i,j)$-th element in matrix $\boldsymbol{B}$ is only perturbed when there exists an electrical line connecting nodes $i$ and $j$. The unperturbed elements in matrix $\boldsymbol{B}$ are marked as NaN (grey color) in Fig. 16. From Fig. 16, the most vulnerable electrical lines for both control schemes can be directly identified, which are line (1,5) for O-NAPC and line (3,6) for A-NAPC. It should also be noted that line (4,6) is the next vulnerable line for A-NAPC. The parameter changes of these electrical lines have the most significant impact on the system stability, implying that the maintenance or monitoring of these lines needs to be enhanced.

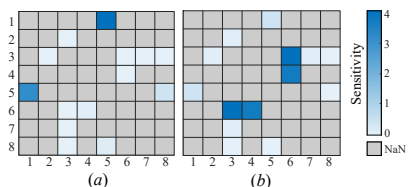

Fig. 16 The vulnerability analysis on the perturbations of matrix $\boldsymbol{B}$: (*a*) O-NAPC control scheme, (*b*) A-NAPC control scheme.

The vulnerability analysis of the two control schemes on the perturbations of matrix $\boldsymbol{L}_W$ is shown in Fig. 17. Different from matrix $\boldsymbol{B}$, every element of matrix $\boldsymbol{L}_W$ is perturbed to imitate the penetrative influence of cyber attacks. The most vulnerable weakness for both control schemes is (7,7), indicating that, with the same attacking resources, cyber attacks on this node would achieve the most significant effect. The high sensitivity of diagonal elements means that this node is a hub node in the communication network. Attacking this node will influence the information exchange among other nodes in the communication network to the largest extent. From the view of matrix $\boldsymbol{L}_W$, the element with the highest sensitivity indicates that with the least perturbation of this element, the positive semi-definite property of matrix $\boldsymbol{L}_W$ can be sabotaged. Note that the vulnerabilities of elements (2,7) and (7,2) are different for the O-NAPC control scheme. This demonstrates that the directed communication links between two nodes can have different impacts on system stability. However, the vulnerabilities of elements (2,7) and (7,2) for the A-NAPC control scheme are the same, because the amplifier restores the symmetric communication process, i.e., $\boldsymbol{D}_p \cdot \boldsymbol{L}_W \cdot \boldsymbol{D}_p$.

The available capacities of individual nodes can be disturbed due to electrical faults, voltage deviation, or end-user behavior. The vulnerability analysis of the two control schemes on the perturbations of matrix $\boldsymbol{D}_p$ is shown in Fig. 18. Node 1 for O-NAPC and node 6 for A-NAPC can be clearly identified as the most important DER in the system, emphasizing the necessity to further enhance these DERs. Moreover, the perturbations of matrix $\boldsymbol{D}_p$ have a higher sensitivity than matrices $\boldsymbol{B}$ and $\boldsymbol{L}_W$, which shows that the capacities of controllable DERs have the most significant influence on the system stability.

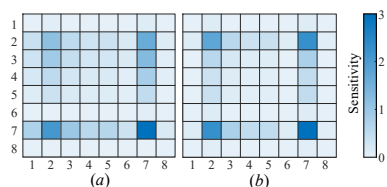

Fig. 17 The vulnerability analysis on the perturbations of matrix $\boldsymbol{L}_W$: (*a*) O-NAPC control scheme, (*b*) A-NAPC control scheme.

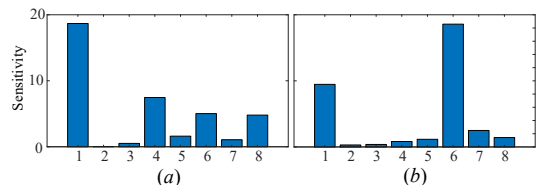

Fig. 18 The vulnerability analysis on the perturbations of matrix $\boldsymbol{D}_p$: (*a*) O-NAPC control scheme, (*b*) A-NAPC control scheme.

## V. Conclusion

This paper, for the first time, provides a thorough theoretical proof of the asymptotic stability of O-NAPC and A-NAPC based droop-free controlled microgrids. Both O-NAPC and A-NAPC control schemes are theoretically proved to be asymptotically stable by testifying that all effective eigenvalues have negative real parts. Both control schemes can also achieve the same steady state with controllable DERs providing power outputs in proportion to their available capacities. Simulation results show that the convergence speeds of these two control schemes depend on their dominant poles, which are influenced by the average admittance of the electrical network, the sparseness of the communication network, and the average available capacity of controllable DERs. The sensitivity analysis further identifies that O-NAPC /A-NAPC is more suitable for microgrids with controllable DERs of larger/smaller available capacities for achieving faster convergence performance. The vulnerability analysis can effectively identify the weak components in the system and set the guidelines for further enhancement.

Future research could extend the proposed work to further theoretically study the asymptotic stability of NAPC-based droop-free control with asymmetric communication networks for its application in practical microgrids. Moreover, other communication issues, such as the impact of communication delay on the system stability and convergence speed, will also be covered in our future research directions.

APPENDIX

## A.1 Adaptability to Generalized Situations with Non-Zero R/X Ratios

Let $\varphi = \arctan(R/X)$ represent the non-zero $R/X$ ratio. Because the nearby nodal phase angles under the normal operation states are close to each other, we have $\sin\theta_{ij} \approx \theta_{ij}$ and $\cos\theta_{ij} \approx 1$. With this, the linearized power flow equation can be represented as in (26), where $|V|(t)$ and $q_{net}(t)$ are vectors of nodal voltage magnitudes and net reactive power, and $G$ is the conductance matrix. Clearly, active and reactive power are coupled due to the non-zero blocks $B$ and $G$. Thus, the droop-free control law (8) with NAPC that only adjusts $\theta(t)$ will not be able to achieve frequency synchronization and active power sharing goals.

$$\begin{bmatrix} p_{net}(t) \\ q_{net}(t) \end{bmatrix} = \begin{bmatrix} B & -G \\ G & B \end{bmatrix} \cdot \begin{bmatrix} \theta(t) \\ |V|(t) \end{bmatrix} \quad (26)$$

Frame transformation can be used to decouple the interaction of active and reactive power flows [21]. A frame transformation matrix $T$ is added in the original droop-free controller (8) to derive (27), where $I_2$ is the second-order identity matrix, $\otimes$ represents the Kronecker product, and $q_c(t)$ is the nodal reactive power vector of controllable distributed energy resources. The frame transformation matrix $T$ is calculated as in (28) using $R/X$ ratios $\varphi = \arctan(R/X)$.

$$\begin{bmatrix} \theta(t) \\ |V|(t) \end{bmatrix} = \begin{bmatrix} \omega_n \cdot \mathbf{1}_N \\ \mathbf{0}_N \end{bmatrix} - h \cdot T \cdot \left( I_2 \otimes L_W \cdot D_p \right) \cdot \begin{bmatrix} p_c(t) \\ q_c(t) \end{bmatrix} \quad (27)$$

$$T = \begin{bmatrix} \cos\varphi & -\sin\varphi \\ \sin\varphi & \cos\varphi \end{bmatrix} \otimes I_N \quad (28)$$

With this, the system matrix $A_{RX}$ of the droop-free controlled microgrid after the frame transformation is given as in (29). Because $B\sin\varphi = -G\cos\varphi$, (29) can be simplified to (30), where $B_{RX} = B\cos\varphi - G\sin\varphi$ is a symmetric Laplacian matrix.

$$A_{RX} = -h \cdot \begin{bmatrix} B & -G \\ G & B \end{bmatrix} \cdot T \cdot \left( I_2 \otimes L_W \cdot D_p \right) \quad (29)$$

$$= -h \cdot \begin{bmatrix} B\cos\varphi - G\sin\varphi & -B\sin\varphi - G\cos\varphi \\ B\sin\varphi + G\cos\varphi & B\cos\varphi - G\sin\varphi \end{bmatrix} \cdot \left( I_2 \otimes L_W \cdot D_p \right)$$

$$A_{RX} = -h \cdot \begin{bmatrix} B_{RX} \cdot L_W \cdot D_p & \\ & B_{RX} \cdot L_W \cdot D_p \end{bmatrix} \quad (30)$$

Thus, active power and reactive power are decoupled, and the system matrix of the active power state space is $-h \cdot B_{RX} \cdot L_{asym} \cdot D_p$, which has the same form as the proposed droop-free control with NAPC shown in (10). In other words, the proposed droop-free control can be applied to microgrids with non-zero $R/X$ ratios, where $I_N$ is the Nth-order identity matrix.

## A.2 Adaptability to Including Filter Dynamics

Filters are often deployed between system measurements and controllers to obtain clean signals. The filter dynamics can be formulated as in (31), where $\tau$ is the filter time constant and $p_{c,f}(t)$ is the filtered active power signal.

$$\dot{p}_{c,f}(t) = -\frac{1}{\tau} p_{c,f}(t) + \frac{1}{\tau} p_c(t) \quad (31)$$

By including filter dynamics, the input of the controller is changed from $p_c(t)$ to $p_{c,f}(t)$, expanding the system size from $N$ to $2N$. The system state space equation is rewritten as in (32).

$$\dot{p}_f(t) = A_f \cdot p_f(t) \quad (32)$$

where $p_f(t) = [p_{c1}(t), p_{c1,f}(t), p_{c2}(t), p_{c2,f}(t), \ldots, p_{cN}(t), p_{cN,f}(t)]^T$ is the vector consisting of active power, and $A_f$ is the system matrix, which is given as in (33) and (34).

$$A_f = I_N \otimes A_{f,1} + B \cdot L_W \cdot D_p \otimes A_{f,2} \quad (33)$$

$$A_{f,1} = \begin{bmatrix} 0 & 0 \\ \dfrac{1}{\tau} & -\dfrac{1}{\tau} \end{bmatrix}, \quad A_{f,2} = \begin{bmatrix} 0 & -h \\ 0 & 0 \end{bmatrix} \quad (34)$$

According to Schur's triangulation theorem, there exists a unitary matrix $U$ that can transform $B \cdot L_W \cdot D_p$ to an upper-triangular matrix with the same spectrum. Applying the same unitary matrix $U$ on matrix $A_f$, we can obtain $A_{fu}$ as in (35).

$$A_{fu} = \left( U^{-1} \otimes I_2 \right) \cdot A_f \cdot \left( U \otimes I_2 \right)$$
$$= \left( U^{-1} \otimes I_2 \right) \cdot \left( I_N \otimes A_{f,1} \right) \cdot \left( U \otimes I_2 \right) \quad (35)$$
$$+ \left( U^{-1} \otimes I_2 \right) \cdot \left( B \cdot L_W \cdot D_p \otimes A_{f,2} \right) \cdot \left( U \otimes I_2 \right)$$

Because $I_N \otimes A_{f,1}$ is a block diagonal matrix and $B \cdot L_W \cdot D_p \otimes A_{f,2}$ is an upper triangular matrix, $A_{fu}$ is an upper-triangular matrix whose eigenvalues are the union of eigenvalues for its block diagonal submatrices. The block diagonal submatrices of $A_{fu}$ can be represented as in (36), where $s_i$ is the $i$-th eigenvalue of matrix $-h \cdot B \cdot L_W \cdot D_p$. The eigenvalues of the state matrix $A_{fu}$ can be derived as in (37). Since time constant $\tau$ is positive and eigenvalues $s_i$ are non-positive, $\lambda_{Afu,i}$ is non-positive. By applying similar proof of *Proposition* 1, we can validate that the filter-included system has $N$-1 effective negative eigenvalues and thus remains asymptotically stable.

$$A_{fu,dia} = \begin{bmatrix} 0 & s_i \\ \dfrac{1}{\tau} & -\dfrac{1}{\tau} \end{bmatrix} \quad (36)$$

$$\lambda_{Afu,i} = \frac{-1 \pm \sqrt{1 + 4\tau s_i}}{2\tau} \quad (37)$$

## A.3 Adaptability to Generalized Situations with Only Certain Nodes Being Associated with Controllable DERs

Considering an $N$-node electrical network, $N_c$ nodes are connected with controllable DERs, and $N > N_c$. The admittance matrix of the electrical network can be shown as in (38), where the nodes are ordered in a way that the first $N_c$ nodes are those connected with controllable DERs. Thus, subblock $Y_{11}$ corresponds to the $N_c$ nodes connected with controllable DERs, subblock $Y_{22}$ corresponds to the $(N$-$N_c)$ nodes without controllable DERs, and $Y_{12}^T = Y_{21}$.

$$Y = \begin{bmatrix} Y_{11} & Y_{12} \\ Y_{21} & Y_{22} \end{bmatrix} \quad (38)$$

The net current injection for each of the $(N$-$N_c)$ nodes without controllable DERs is zero. Thus, the voltage equation of these nodes is shown as in (39). $V_c$ is the nodal voltage vector of the $N_c$ nodes connected with controllable DERs, and $V_u$ is the nodal voltage vector of the $(N$-$N_c)$ nodes without controllable DERs.

$$Y_{21} \cdot V_c + Y_{22} \cdot V_u = 0 \quad (39)$$

Similarly, the voltage equation of the $N_c$ nodes connected with controllable DERs is shown as in (40), where vector $I_c$ is the net current injections of these nodes.

$$Y_{11} \cdot V_c + Y_{12} \cdot V_u = I_c \quad (40)$$

According to the properties of the admittance matrix $Y$, the submatrix $Y_{22}$ is a positive definite matrix and is invertible [28]. Thus, we have $V_u = -Y_{22}^{-1} \cdot Y_{21} \cdot V_c$ from (39). Substituting $V_u$ into (40) derives (41).

$$\left(Y_{11} - Y_{12} \cdot Y_{22}^{-1} \cdot Y_{21}\right) \cdot V_c = I_c \tag{41}$$

Therefore, we have the equivalent admittance matrix $Y_{eq} = Y_{11} - Y_{12} \cdot Y_{22}^{-1} \cdot Y_{21}$, representing an equivalent electrical network with $N_c$ nodes and each node having a controllable DER.

Note that $Y_{eq}$ is the Schur complement of $Y$, as $Y$ is a semi-positive-definite matrix, $Y_{eq}$ is also a semi-positive-definite matrix [29]. Thus, using $Y_{eq}$ to replace the original $Y$ will not change the properties of the system matrix (10) and (13) in this paper, and their asymptotic stability will not be affected.